\documentclass[twocolumn]{aa}
\usepackage{graphicx,epsfig}
\usepackage{txfonts}
\newcommand{\kms}{km~s$^{-1}$}

\newcommand{\cm}{cm$^{-2}$}

\newcommand{\NHI}{N$_{\rm HI}$}

\newcommand{\lya}{Lyman-$\alpha\ $}
\newcommand{\acc}{~cm$^{-2}$}
\begin{document}
\title{The strange case of a sub-DLA with very little HI}

\author{Nissim Kanekar
          \inst{1}
          \and
          Jayaram N. Chengalur \inst{2}
          }
\authorrunning{Kanekar \& Chengalur }
\offprints{J. N. Chengalur}
\institute{Kapteyn Astronomical Institute, University of Groningen, 
	   Post Bag 800, 9700 AV Groningen, The Netherlands \\
           \email{nissim@astro.rug.nl}
	   \and
	   National Centre for Radio Astrophysics, Post Bag 3, 
	   Ganeshkhind Pune - 411 007, India \\
           \email{chengalu@ncra.tifr.res.in}
             }

\date{Received ; accepted }

\abstract{ We report a deep search for HI 21cm emission from 
the $z= 0.00632$ sub-DLA toward PG1216+069 with the Giant 
Metrewave Radio Telescope. No emission was detected and 
our $5\sigma$ upper limit on the mass of any associated galaxy 
is  $M_{\rm HI} \lesssim 10^7 M_\odot$, nearly 3
orders of magnitude less than M$_{\rm HI}^\star$. The 
$z \sim 0.006$ absorber is thus the most extreme known 
deviation from the standard paradigm in which high column 
density quasar absorption lines arise in the disks of 
gas-rich galaxies.
   }

   \maketitle
%

\section{Introduction}

      Neutral atomic gas clouds at high redshift can currently be detected 
only via the absorption lines they produce if they happen to lie
along the line of sight to an even more distant quasar. Although 
present technology allows one to detect \lya absorption lines from 
clouds with as low an HI column density  as \NHI~$\sim 10^{13}$\acc, 
the bulk of the neutral gas at $z \sim 3$ turns out to be in extremely 
rare systems, with column densities higher by several orders of magnitude, 
viz. \NHI~$\gtrsim 10^{20}$\acc. At such large \NHI, 
the optical depth is substantial even in the Lorentzian wings of the 
Lyman-$\alpha$ profile; these systems are called damped 
Lyman$-\alpha$ absorbers or DLAs. 

        In the local universe, the disks of spiral galaxies 
have characteristic HI column densities $\gtrsim 10^{20}$\acc~over a 
large spatial scale. High $z$ DLAs are thus natural candidates 
for the precursors of today's large galaxies (e.g. \cite{wolfe88}). 
The nature of high redshift DLAs and their evolution with redshift 
have hence been of much interest to astronomers studying
galaxy formation and evolution. Unfortunately, cosmological dimming 
makes it impossible to directly image high redshift DLAs; the information 
obtainable is hence limited to the gas that lies along the narrow 
pencil beam that is illuminated by the background quasar. It is 
primarily for this reason that, despite decades of study, the nature 
of high redshift damped absorbers remains an unsettled issue, with 
models ranging from large, rotating disks (e.g. \cite{prochaska97})
to small, merging proto-galaxies (e.g. \cite{haehnelt98}). Detailed 
absorption studies with 10-metre class optical telescopes have 
established that current samples of high redshift DLAs typically 
have low metal abundances ($\lesssim 0.1$~solar; \cite{pettini97})
and that these abundances evolve slowly, if at all, with redshift
(e.g. \cite{prochaska03,kulkarni04}). This is somewhat surprising 
if the absorbers are indeed typically the precursors of large 
galaxies like the Milky Way.

    Low redshift ($z \lesssim 0.5$) DLAs offer the possibility of 
direct identification of the absorbing galaxy by optical imaging, 
(e.g. \cite{lebrun97,rao03,chen03}). Since the HI mass function is 
dominated by bright galaxies (e.g. \cite{zwaan97}), one might 
{\it a priori} expect that low $z$ DLAs should primarily be associated 
with large spirals. Interestingly enough, however, it has been found 
that a wide variety of galaxy types are responsible for damped 
absorption at low redshifts (e.g. \cite{rao03,bowen01}), with no 
particular type dominating the current (admittedly rather small) 
low $z$ sample. One possible explanation is that, although the 
total HI mass is dominated by large spiral galaxies, the HI 
cross-section at the $\sim 10^{20}$\acc~ threshold varies across 
galaxy type in such a way as to yield an even distribution of 
absorbers across a range of galaxy luminosities (\cite{zwaan02}). 
On the other hand, it is also possible that, although the optical 
counterparts to these low redshift DLAs are faint, they have 
unusually large HI envelopes and thus, a large {\it total} mass. 
Such a preferential selection might well occur since absorption 
surveys are biased toward objects with a large HI cross-section. 
Unfortunately, it is very difficult to observationally test the 
latter scenario as this requires detection of the HI 21-cm line 
and current radio instrumentation limits searches for 21-cm 
emission from galaxy-sized objects to the very nearby 
universe, $z \lesssim 0.1$. Almost no DLAs are known at these 
low redshifts and hence, searches for 21-cm emission have 
been carried out in only three absorbers; in 
all cases, the HI mass has been found to be lower than 
$M_{\rm HI}^\star$, the typical mass of local bright spirals 
(\cite{lane00,kanekar01,bowen01,chengalur02}). Identifying such 
$z\lesssim0.1$ DLAs is of much importance as it is only 
these systems that can be followed up in detail in all wavebands, 
to estimate the total mass (as opposed to the luminous mass). We 
note, in passing, that such DLAs {\it cannot} be identified (except 
by accident) using the MgII selection criterion of Rao \& 
Turnshek~(2000), as the MgII lines are redshifted into optical 
wavebands only for redshifts $z > 0.1$.

Recently, Tripp et al. (2004) reported the discovery 
of an extremely low redshift ($z = 0.00632$) absorption system toward 
the quasar PG1216+069, with a high HI column density, 
\NHI$ \sim 2\times10^{19}$~\acc. While the 
absorption profile shows clear damping wings, its column density is 
somewhat smaller than the cut-off of $2 \times 10^{20}$\acc~that 
has traditionally been used to define DLAs (e.g. \cite{wolfe86}). 
This system is hence classified as a sub-DLA. However, it should 
be emphasized that the column density threshold used to define DLAs 
is entirely arbitrary and that the absorber properties do not show any 
sharp qualitative differences here.
 
Interestingly enough, although the new absorber is at a very low 
redshift, it has an extremely low metallicity, $\sim 1/40$~Solar 
(\cite{tripp04}), one of the lowest measured in the gas phase in 
the nearby universe and, in fact, similar to that of high $z$ 
DLAs (e.g. \cite{prochaska03}). No optical counterpart can be 
seen in the Hubble Space Telescope (HST) image of the field 
(\cite{tripp04}); the nearest $L_*$ galaxy has a projected 
separation of $\sim 252 \: h^{-1}_{71}$~kpc\footnote{We assume 
$\Omega_m = 0.27$, $\Omega_\Lambda = 0.73$ and $H_0 = 
71$~km~s$^{-1}$~Mpc$^{-1}$ throughout this paper.}. We report
here the results of a search for HI 21-cm emission from the 
$z \sim 0.006$ sub-DLA, using the Giant Metrewave Radio 
Telescope (GMRT; \cite{swarup91}). 

\section{Observations, data reduction and results}

     The GMRT observations were conducted on the 28th and 30th 
of August 2004. The total on-source time was $\sim 8$~hours, with 
29 antennas. The observing bandwidth of 2~MHz, centred at 1411.5~MHz, 
was divided into 128 spectral channels, yielding a spectral 
resolution of $\sim 15.6$~kHz. The total velocity coverage was thus 
$\sim 425$~\kms, with a resolution of $\sim 3.3$~\kms. Flux and 
bandpass calibration were done using short scans on 3C147 and 3C286 at 
the start and end of each 
observing run, while the compact source 1150$-$003 was used 
for phase calibration. The flux density of 1150$-$003 was 
measured to be $2.7 \pm 0.1$~Jy.
    
     Data reduction was carried out using standard tasks in ``classic'' 
AIPS. Visibilities from the different days were calibrated separately 
and then combined together using the AIPS task DBCON; the combined 
data were used to make the final images, with the task IMAGR. 
We note that the GMRT has a hybrid configuration, with fourteen antennas 
in a central array (the ``central square'') and the remaining sixteen 
distributed in the three arms of a ``Y'' (\cite{swarup91}). The central 
square antennas yield baselines of $\lesssim 1$~km (i.e. U-V coverage out 
to $\sim 5 k\lambda$ at 1411.5~MHz), while longer baselines (up to a 
maximum length of $\sim 25$~km, or a U-V coverage to $\sim 120 k\lambda$ 
at 1411.5~MHz) are obtained with the arm antennas. Continuum images and 
spectral cubes were hence made with a variety of U-V ranges and tapers,
allowing a search for HI and continuum emission at various spatial 
resolutions ranging from $\sim 3''$ to $\sim 40''$.  
The continuum emission was subtracted using two independent approaches: (1)
in the U-V plane, using the task UVSUB and (2) in the image plane, 
using the task IMLIN. The two strategies gave very similar results; the 
numbers quoted below are for the U-V plane approach.

Finally, the GMRT does not do on-line Doppler tracking, implying that 
any required Doppler shifts must be applied to the data off-line. 
In the present case, however, it was not necessary to apply a 
differential Doppler shift to the data of the two observing runs 
since this shift was found to be small, relative to the channel 
separation. No off-line Doppler corrections were hence applied 
while making the image cubes.

\begin{table}[t!]
\caption{RMS noise levels and $5\sigma$ HI mass limits}
\begin{tabular}{ccccc}
\hline \hline
$\theta_{HPBW}$ & RMS & r$_{\rm HPBW}$ & M$_{\rm HI}$\\
bmaj$'' \times $ bmin$''$ & mJy/Bm & $h^{-1}_{71}$ kpc & $10^6$ $M_\odot$\\
\hline
&&& \\
$39.4 \times 37.7$ & 0.62 & 5.0 & 11\\
$26.1 \times 23.1$ & 0.52 & 3.3 & 8.8\\
$12.3 \times 9.9$  & 0.45 & 1.6 & 7.7\\
$2.9  \times 2.5$  & 0.30 & 0.4 & 5.0\\
&&& \\
\hline \hline
\end{tabular}
\label{tab:res}
\end{table}

The central ninety channels of the band were averaged together to 
map the continuum, at various spatial resolutions. The low resolution 
($\sim 39.4'' \times 37.7''$) image shows good agreement with the NRAO 
VLA Sky Survey map of the same region (\cite{condon98}). Our somewhat 
higher sensitivity allows us to detect faint emission from the quasar 
itself; we measure a flux density of $\sim 2.8 \pm 0.3$~mJy. The 
peak flux density in the image is $\sim 14.6 \pm 0.3$~mJy.

The data cubes were examined for line emission at a variety of 
spectral resolutions; in all cases, no significant line emission was 
found in the vicinity of the quasar line-of-sight. Besides a visual 
inspection, the AIPS task SERCH was used to search for line emission 
in the different cubes. No statistically significant emission features 
were detected in the cube, except from the galaxy VCC~297; this system 
is discussed later. The highest resolution ($2.9''  \times 2.5''$) 
cube did show a weak ($3.3 \sigma$) emission feature at the quasar 
location (albeit $\sim 70$~\kms~ away from the optical redshift); if 
real, it corresponds to an HI mass of $4 \times 10^6 M_\odot$. However, 
this is not very believable as the feature is spatially unresolved 
and seen only in a single velocity channel, after smoothing the data 
to a resolution of 13~\kms. Further, if real, this concentration of gas 
would imply an HI column density $\gtrsim 10^{21}$~\acc, (i.e. 
considerably larger than the column density of the sub-DLA) and 
it would be surprising that no associated metal lines were found 
at this velocity.

Table~\ref{tab:res} summarizes our results, for a representative 
selection of spatial resolutions; the data have been smoothed 
to a velocity resolution of 20~\kms~in all cases, a fairly typical 
velocity width for a small galaxy. The different 
columns in the table are: (1)~the half-power beam width (HPBW) of the 
synthesized beam (i.e. the spatial resolution, in arcsecs.), (2)~the 
RMS noise (in mJy/Bm) at this spatial resolution, (3)~the physical 
distance at $z = 0.00632$ (in kpc), corresponding to the synthesized 
HPBW, and (4)~the $5\sigma$ upper limit on the HI mass of the $z \sim 0.006$ 
sub-DLA at this spatial resolution (in units of $10^6 M_\odot$), assuming that 
the HI profile has a top-hat shape, with a velocity width of 20~\kms. 
Note that lower spatial resolutions are obtained by weighting down 
data from the more distant antennas; this implies that the mass limit 
{\it improves} at higher spatial resolution.

\section{Discussion}

The $5\sigma$ upper limits on the HI mass of the galaxy listed in 
the last column of Table~\ref{tab:res} are extremely small, nearly
3 orders of magnitude lower than $M_{\rm HI}^\star$. This is 
quite remarkable, given the high HI column density estimated from
the Lyman-$\alpha$ line. Of course, it should be pointed out that 
the mass limits quoted in the table are based on the assumption 
that the absorber is  entirely contained within the synthesized beam.
It is, in principle, possible that the HI emission is actually 
spread over a larger angular scale and is not detected in our 
interferometric observations either because the flux per synthesized 
beam falls below our detection threshold or because the flux is resolved
out or due to a combination of both effects. We initially consider this
possibility, before discussing the implications  of our results 
for the nature of the $z = 0.00632$ sub-DLA.

If the absorbing galaxy were far larger than the GMRT synthesized beam,
the mass limits of Table~\ref{tab:res} would refer only 
to the HI mass within the beam and not to that of the entire galaxy.
For example, if the galaxy were a face-on disk of diameter 10~kpc, 
twice our coarsest resolution, the upper limit to its {\it total} 
HI mass would be a factor of $\sim 4$ larger than that listed in 
the first line of Table~\ref{tab:res}, i.e. $M_{\rm HI} \lesssim 
4 \times 10^7 M_\odot$. However, dwarf irregular galaxies with 
low HI masses ($\lesssim 10^8 M_\odot$) have fairly 
small HI diameters (e.g. \cite{stil02a,stil02b,begum03}). For example, 
all ten dwarf irregulars with $M_{\rm HI} \lesssim 10^8 M_\odot$ 
in the sample of Stil \& Israel (2002a, 2002b) have size~$\lesssim 5$~kpc, 
our coarsest resolution. In fact, of the 27~dwarfs in 
the Stil \& Israel sample with estimates of HI extent, the only
galaxies that have physical sizes larger than 11~kpc are the five 
systems with $M_{\rm HI} \gtrsim 6.5 \times 10^8 M_\odot$ (i.e.
considerably larger than the upper limit in Table~\ref{tab:res}).
Further, if we assume that \NHI~$\sim 2 \times 10^{19}$\acc~ 
throughout the absorber, the total HI mass in an area 5~kpc in 
diameter is $\sim 3 \times 10^6 M_\odot$, comparable to our detection 
threshold. The latter is a severe under-estimate of the true HI 
mass, if the absorption arises in a normal galaxy, since large 
fractions of even dwarf galaxy disks have \NHI~$\gtrsim 10^{20}$\acc. It is 
thus highly unlikely that the absorption arises in a dwarf galaxy 
whose HI disk is so extended that the flux per synthesized beam 
falls below our detection threshold.  Similarly, the second scenario, 
that the flux is resolved out because it arises from a highly 
uniform HI distribution, is also rather improbable as this would 
require a very small velocity gradient over the entire spatial 
extent of the gas, for emission to not be detected even at the 
highest velocity resolutions. 

\begin{figure}[t!]
\centering
\epsfig{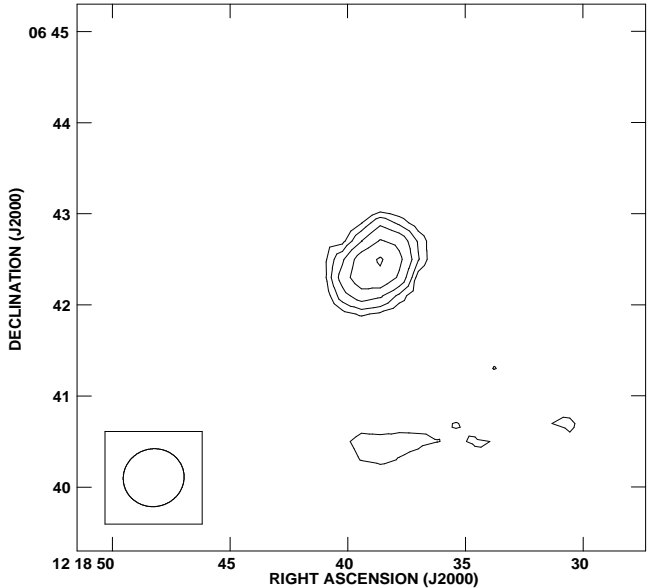}
\caption{Integrated HI 21-cm emission profile toward VCC~297, 
made from the $39'' \times 34''$ resolution image cube. 
The contours are at 0.08,0.12,0.17,0.25 and 0.36 Jy/Bm~\kms. }
\label{fig:vcc297}
\end{figure}

    The nearest known HI-rich galaxy to the sub-DLA is VCC~297, 
which has $L = 0.25 L_\star$ (\cite{impey99}), $M_{\rm HI}=2.7\times10^8 
M_\odot$ and a velocity width W$_{50}$ of 145~\kms~ (\cite{giovanelli97}); 
this is $\sim 87 \: h^{-1}_{71}$~kpc away from the QSO sightline. VCC~297 lies 
at the edge of our field of view, where imaging is particularly difficult 
because of the asymmetry of the edges of the GMRT primary beam. The 
emission velocity of VCC~297 also lies at the edge of the observing 
band where the sensitivity is lower; further, there appears to be some 
low level interference at these edge channels (note that none of 
these problems are present at the centre of our field of 
view and observing band, where the emission from the sub-DLA 
would arise). Next, the optical extent of VCC~297 is a 
factor of $\sim 3$ larger than our synthesized beam; its HI extent 
is likely to be at least this large, reducing the flux per synthesized 
beam. Despite these problems, HI emission from 
VCC~297 is reliably detected in the spectral cubes 
(see Fig.~\ref{fig:vcc297}). The emission was readily 
apparent to the eye and was also detected by the automatic 
search algorithm SERCH, at a signal-to-noise ratio of $\sim 12$. 
Finally, as a concrete counter-example, a dwarf galaxy at 
$z=0.0051$ with $M_{\rm HI} \sim 2\times 10^7 M_\odot$ was 
serendipitously discovered in a different GMRT observation, 
with essentially identical observational setup and RMS noise 
(\cite{chengalur04}).  From all of the above, we conclude that 
it is highly unlikely that the sub-DLA arises in a 
galaxy with HI mass substantially larger than the limits
quoted in Table~\ref{tab:res}. Further, if one assumes a uniform 
HI column density of $2 \times 10^{19}$~\cm~ throughout the 
absorber (and a disk geometry), our HI mass limit of $\sim 10^7 M_\odot$ 
implies an upper limit of $\sim 9$~kpc to the size of any associated 
galaxy. It is still not impossible, of course, that the sub-DLA 
arises in a more massive, highly extended, smooth HI cloud, but this 
would require it to be completely unlike any known low redshift galaxy.

As discussed in detail by Tripp~et~al.~(2004), the $z \sim 0.006$ 
sub-DLA is highly unusual in having a very low metallicity 
([O/H]~$= -1.6^{+0.09}_{-0.11}$) despite being at low redshift and 
lying in a region of high local galaxy density (the outskirts of the 
Virgo cluster). The sub-DLA also shows an under-abundance of nitrogen 
and an over-abundance of iron (implying a lack of dust) which also argues
in favour of a system that is at an early stage of chemical evolution 
(\cite{tripp04}). These abundance patterns are very different from 
those characteristic of local $L_\star$ spirals, and are consistent 
with the absence of a luminous optical counterpart to the sub-DLA 
in the HST image. Of course, the lack of a detected optical counterpart
does not entirely exclude a scenario in which the sub-DLA arises 
in a massive galaxy, since such 
a system might well be hidden beneath the point spread function of 
either the QSO itself or a nearby foreground star located near the quasar 
(\cite{tripp04}); alternatively, as discussed in the introduction, 
the absorber might be anomalously gas-rich for its luminosity. However, 
the stringent upper limit on the HI mass of the absorber provided by 
our observations indeed rules out this possibility.

The abundance patterns and low metallicity of the $z \sim 0.006$ 
absorber are consistent with an origin in (i)~a low metallicity, 
gas-rich, blue compact dwarf (BCD) like I~Zw~18 or 
SBS0335$-$052, (ii)~a high velocity cloud (HVC), 
(iii)~a dwarf spheroidal galaxy or (iv)~a dark mini-halo left over from the 
epoch of reionization (see Tripp~et~al.~(2004) and references therein). 
However, the mass limits in 
Table~\ref{tab:res} are significantly lower than HI masses typical 
of low metallicity BCDs ($\gtrsim 10^8 M_\odot$; 
\cite{vanzee98,pustilnik01}). It 
is also unlikely that the sub-DLA is associated with an HVC, 
given that the nearest $L_\star$ galaxy (NGC4260) 
is $\sim 252 \:  h^{-1}_{71}$~kpc away (in projection) and HVCs 
have not been detected at such large distances from their 
parent galaxies, despite deep searches (e.g. \cite{pisano04}). 
In fact, for M31 (the only galaxy that has been sensitively searched for 
large-scale diffuse HI emission), the typical HI column density 
falls to values smaller than $10^{15}$\acc~ 
by the time one reaches galacto-centric distances of 200~kpc 
(\cite{braun04}). Similarly, Miller \& Bregman~(2004) place a 
limit of $\sim 25$~kpc on the distance of HVCs from their 
parent galaxy; this implies that the absorption is also 
unlikely to arise from an HVC associated with the fainter 
galaxy VCC~297, which is $\sim 87 \: h^{-1}_{71}$~kpc away. 
The most likely counterpart is hence an extreme dwarf galaxy 
(or a dark mini-halo), which is surprising, given the relatively 
small size of these systems. How likely is it that such a system would 
show up in an unbiased search for absorption? 
While there are no existing measurements of the typical size of 
faint dwarf galaxies at HI column densities of $2\times 10^{19}$~\acc, 
a linear extrapolation of the results of Zwaan~et~al.~(2002), 
for systems with higher mass and column density, (and further 
assuming the faint end slope of the HI mass function to be $-1.2$) 
indicates that the HI cross-section offered by $\sim 10^7$~M$_{\odot}$ 
galaxies is significantly more than an order of magnitude smaller than that 
of M$_{\rm HI}^\star$ galaxies. The implied low probability of 
a sub-DLA arising in the former class of systems may suggest that 
our understanding of the HI mass function is incomplete at the low mass 
end.

In summary, contrary to the {\it a priori} expectation that 
systems found via absorption line searches should be biased toward 
galaxies with large HI mass, our upper limit to the total HI content 
of the $z \sim 0.006$ sub-DLA is $\sim$ three orders of magnitude 
lower than $M_{\rm HI}^\star$. This is by far the smallest HI 
mass limit placed on gas associated with DLAs or sub-DLAs. 
The $z= 0.00632$ sub-DLA toward PG1216+069 thus appears to be 
the most extreme known deviation from the standard paradigm for 
high column density absorbers.

\begin{acknowledgements}
The data presented in this paper were obtained using the GMRT,
which is operated by the National Centre for Radio Astrophysics 
of the Tata Institute of Fundamental Research. We thank the referee,
Martin Zwaan, for useful comments on the manuscript.
\end{acknowledgements}


\begin{thebibliography}{}
\bibitem[Bowen et al. 2001]{bowen01} Bowen D.~V., Huchtmeier W., Brinks E., 
  	Tripp T.~M. \& Jenkins E.~B., 2001, A\&A, 372, 820 
\bibitem[le Brun et al. 1997]{lebrun97} le Brun V., Bergeron J., 
	Boiss\'{e} P. \&    Deharveng J-M., 1997, A\&A, 321, 733
\bibitem[Begum, Chengalur \& Hopp 2003]{begum03} Begum A., Chengalur J.~N. 
	\& Hopp U., 2003, NewA, 8, 267 
\bibitem[Braun \& Thilker 2004]{braun04} Braun R. \& Thilker D. A.,
        2004, A\&A, 417, 421
\bibitem[Chen \& Lanzetta 2003]{chen03} Chen H-W. \& Lanzetta K.~M., 2003, 
	ApJ, 597, 706
\bibitem[Chengalur \& Kanekar 2002]{chengalur02} Chengalur J.~N. \& 
	Kanekar N., 2002, A\&A, 388, 383 
\bibitem[Chengalur et al. 2004]{chengalur04} Chengalur J.~N. et al., 2004, 
	in preparation
\bibitem[Condon et al. 1998]{condon98} Condon J. J., Cotton W. D., 
	Greisen E. W., Yin Q. F., Perley R. A., Taylor G. B. \& 
	Broderick J. J., 1998, AJ, 115, 1693
\bibitem[Giovanelli, Avera \& Karachentsev 1997]{giovanelli97} Giovanelli R., 
	Avera E. \& Karachentsev I. D., 1997, AJ, 114, 122
\bibitem[Haehnelt, Steinmetz \& Rauch 1998]{haehnelt98} Haehnelt 
	M. G., Steinmetz M. \& Rauch M., 1998, ApJ, 495, 64.
\bibitem[Impey et al. 1999]{impey99} Impey C. D., Petry C. E. \& Flint K. P.
	1999, ApJ, 524, 536
\bibitem[Kanekar et al. 2001]{kanekar01} Kanekar N., Chengalur J.~N., 
	Subrahmanyan R. \& Petitjean P., 2001, A\&A, 367, 46 
\bibitem[Kulkarni et al. 2004]{kulkarni04} Kulkarni V.~P. et al., 2004, 
	accepted to AJ (astro-ph/0409234)
\bibitem[Lane 2000]{lane00} Lane W. M., 2000, Ph.~D. thesis, Univ.~Groningen
\bibitem[Miller \& Bregman 2004]{miller04} Miller E. D. \& Bregman J. N., 
	2004 (astro-ph/0410238)
\bibitem[Pettini et al. 1997]{pettini97} Pettini M., Smith L. J., 
       King D. L. \& Hunstead R. W., 1997, ApJ, 486, 665
\bibitem[Pisano et al. 2004]{pisano04} Pisano D.~J., Barnes D.~G., 
	Gibson B.~K., Staveley-Smith L., Freeman K.~C. \& 
 	Kilborn V.~A., 2004, ApJ, 610, L17 
\bibitem[Prochaska \& Wolfe 1997]{prochaska97} Prochaska J. X. \& 
	Wolfe A. M., 1997, ApJ, 487, 73
\bibitem[Prochaska et al. 2003]{prochaska03} Prochaska J. X., Gawiser E., 
	Wolfe A. M., Castro S. \& Djorgovski S. G. 2003, ApJ, 599, L9
\bibitem[Pustilnik et al. 2001]{pustilnik01} Pustilnik S. A., Brinks E.,
	Thuan T. X., Lipovetsky V. A. \& Izotov Y. I., 2001, AJ, 121, 1413
\bibitem[Rao et al. 2003]{rao03} Rao S. M., Nestor D. B., Turnshek D. A., 
  	Lane W. M., Monier E. M. \& Bergeron J. 2003, ApJ, 595, 94.
\bibitem[Stil \& Israel 2002a]{stil02a} Stil J. M. \& Israel F. P., 2002,
	A\&A, 389, 29
\bibitem[Stil \& Israel 2002b]{stil02b} Stil J. M. \& Israel F. P., 2002,
	A\&A, 389, 42
\bibitem[Swarup et al. 1991]{swarup91} Swarup~G., Ananthakrishnan~S., 
	Kapahi V.~K., Rao A.~P., Subrahmanya C.~R. \& Kulkarni V.~K., 
	1991, Current Science, 60, 95
\bibitem[Tripp et al. 2004]{tripp04} Tripp T. M., Jenkins E.~B., Bowen D.~V.,
	Prochaska J.~X., Aracil B. \& Ganguly R., 2004, submitted to ApJ 
	(astro-ph/0407465)
\bibitem[Wolfe et al. 1986]{wolfe86} Wolfe A. M., Turnshek D. A., Smith H. E. 
	\& Cohen R. D., 1986, ApJS, 61, 249
\bibitem[Wolfe 1988]{wolfe88} Wolfe A. M., 1988, in QSO Absorption 
	Lines: Probing the Universe, J. C. Blades et al. eds., 
	Cambridge University Press
\bibitem[van Zee et al. 1998]{vanzee98} van Zee L., Westpfahl D., 
	Haynes M., \& Salzer J. J., 1998, AJ, 115, 1000 
\bibitem[Zwaan et al. 1997]{zwaan97} Zwaan M. A., Briggs F. H., 
	Sprayberry D. \& Sorar E., 1997, ApJ, 490, 173
\bibitem[Zwaan et al. 2002]{zwaan02} Zwaan M., Briggs F. H. \& 
	Verheijen M. 2002, in ASP Conf. Ser. 254, Extragalactic Gas 
	at Low Redshift, J. Mulchaey \& J. Stocke eds. 
	(San Francisco: ASP), 169
\end{thebibliography}
\end{document}